\documentclass[smallextended]{svjour3}

\usepackage{graphicx}

\usepackage{graphicx,amssymb,bm}

\def\ket#1{|\,#1\,\rangle}
\def\bra#1{\langle\, #1\,|}

\def\proj#1#2{\ket{#1}\bra{#2}}

\newcommand{\beq}{\begin{equation}}
\newcommand{\eeq}{\end{equation}}
\newcommand{\beqa}{\begin{eqnarray}}
\newcommand{\eeqa}{\end{eqnarray}}

\begin{document}

\title{Simultaneous sorting many quDits using different input ports}

\author{Iulia Ghiu}

\institute{I. Ghiu \at University of Bucharest, Faculty of Physics, Centre for Advanced Quantum Physics, PO Box MG-11, R-077125, Bucharest-Magurele, Romania \\
   \email{iulia.ghiu@g.unibuc.ro} }

\date{Received: date / Accepted: date}

\maketitle

\begin{abstract}
Quantum sorter has gained a lot of attention during the last years due to its wide application in quantum information processing and quantum technologies.
A challenging task is the construction of a quantum sorter, which collect many high-dimensional quantum systems, which are simultaneously incident on different input ports of the device.
In this paper we give the definition of the general quantum sorter of multi-level quantum systems. We prove the impossibility of the construction of the perfect quantum sorter, which works for many particles incident on any input port, while keeping their states unmodified. Further we propose an approximate multi-particle multi-input-port quantum sorter, which performs the selection of the particles in a certain output port according to the properties of the initial states, but changing the final states. This method is useful for those situations which require high speed of quantum state sorting. Thus, the information contained in the initial states of the particles is revealed by the click statistics of the detectors situated in each output port.
\end{abstract}


\section{Introduction}

Multi-input multi-output quantum devices have attracted an increasing interest for many applications in quantum technologies during the last years. They have opened or developed branches of research in the fields of quantum information theory and quantum optics: Boson Sampling \cite{Aaronson}, \cite {Molmer-2014}, \cite{Sabin}, Bell multi-port beam splitter \cite{Beige}, quantum Fourier transform \cite{Pan-prl-2017}, generalized interference \cite{Tichy-2014}, \cite{deGuise-2014}, \cite{Tillmann}, \cite{Menssen}, \cite{Khalid}, \cite{Zukowski}, quantum metrology \cite{Spagnolo}, \cite{Humphreys},
\cite{Motes-2015}, \cite{Chaboyer}, entanglement between spatially separate multi-mode quantum systems \cite{Wiseman}, \cite{Hofmann-2017}, \cite{Ghiu-2003}, \cite{Ghiu-2005}, \cite{Luo-2014}. In quantum information protocols, a key step for performing a task is the measurement of the states of the quantum system. Measuring the states is a challenging problem since one has to determine the possible orthogonal states of the quantum system. A possible way of measuring the state can be performed with the help of the so-called quantum sorter, which selects the states in different output ports.

The quantum sorter became of paramount importance for protocols in quantum information processing, as many implementations require the extraction and manipulation of the information stored by quantum systems \cite{Zeilinger-nature-2014}, \cite{Malik-2014}, \cite{Mirhosseini}, \cite{Zeilinger-book}, \cite{Zeilinger-arx-2017}.
Most of the proposals regarding the sorting of quantum systems are related to the properties of the photons: polarization, orbital angular momentum (OAM) \cite{Leach-2002}, \cite{Berkhout}, \cite{Lavery}, \cite{Boyd-nat-com-2013}, \cite{Schulz}, \cite{Wang}, the total angular momentum \cite{Leach-2004}, or radial modes \cite{Zhou}, \cite{Gu-2017}.

A more general quantum sorter was proposed in Ref. \cite{Ionicioiu}, which is universal, i.e. it enables the selection of the states for an arbitrary degree of freedom of a single quDit (D-level quantum system). This device is characterized by D-input and D-output ports. We show in the present work that the universal quantum sorter of Ref. \cite{Ionicioiu} fails in sorting many quantum systems, which are incident on different input ports. In other words, this universal quantum sorter works properly only in the case when all the particles to be sorted are injected through the same (unique) input port.

Simultaneous sorting of many high-dimensional quantum systems, which are incident on different input ports of the device is still an open and  challenging problem. In this work, we introduce the definition of a general multi-input-port quantum sorter
of quDits. The main contribution is written as a theorem, which provides the expression of the unitary operator of the approximate multi-input-port quantum sorter in terms of the well-known quantum gates of quDits.
This sorter can be used to obtain the information of the initial states by measuring the number of particles in each output port. For example, the photocounting measurements were recently employed for
the estimation of the expectation values of different observables \cite{Kovalenko}. Therefore, the measurements of the number of particles play an essential role for extracting different kinds of information or properties of the initial states.

This article is organized as follows. Section 2 introduces the concept of the general quantum sorter of quDits.
The motivation of our work is presented in Sect. 3, where we show that the quantum sorter given in Ref. \cite{Ionicioiu} cannot perform a simultaneously selection of many particles injected through any input port. This sorter is restricted to select the states only in the case when all the particles are incident on a unique input port. This fact leads us to the motivation to construct a multi-input-port quantum sorter. The two-input-port sorter of the polarization of photons is presented in details in Section 4.
In Section 5 we present the classification of the quantum sorters and prove that the perfect multi-input-port quantum sorter of quDits, which keeps the state of the particles unmodified, does not exist.
Further, we focus on the notion of the multi-input-port quantum sorter, that enables the selection of the particles according to their initial properties. But this device performs an approximate sorting since the final states of the particles are changed. Section 6 is devoted to a theorem, which presents a formula of the unitary operator of the multi-input-port quantum sorter expressed in terms of the quantum gates of quDits. By using this theorem, we construct the quantum circuit of this quantum sorter. In addition, we describe the physical implementation of the four-input-port quantum sorter for photons carrying four-dimensional OAM.
Section 7 summarizes our main results of this paper. In Appendix A we present some results regarding the quantum gates for quDits.

\section{Definition of the general quantum sorter of quDits}

In this section, we introduce the definition of a general quantum sorter. Suppose that we have a $D$-level quantum system, whose states are described by the computational basis $\{ \ket{0}$, $\ket{1}$,..., $\ket{D-1}\}$ in the Hilbert space ${\cal H}_{system}$. In addition, one has to consider the states associated with $D$ ports, also denoted by $\ket{j}$, with $j$= 0, 1, ..., $D-1$ acting in the Hilbert space ${\cal H}_{port}$.

Sorting the quantum states of the system consists of a device that collects a given state of the system through a unique output port. This fact requires that the quantum sorter must have $D$ output ports, one associated with each possible state of the quantum system. In the following we will use the notation $\ket{\mbox{system-state}}\ket{\mbox{port-state}}$ for the total state in the Hilbert space ${\cal H}_{system}\otimes {\cal H}_{port}$, i.e. for the state of the quantum system and for the spatial quantum modes, through which the particles travel.

The quantum sorter is a device that selects the particles characterized by the property $\ket{s}$ of the quantum system by collecting them to the same output port-state, which is denoted by $\ket{s}$.

\vspace{0.5cm}

{\bf Definition 1.} Consider two $D$-dimensional Hilbert spaces ${\cal H}_{system}$ and ${\cal H}_{port}$, the first one associated with a quDit system, while the second one to the states of the $D$ ports.
A device acting in the Hilbert space ${\cal H}_{system}\otimes {\cal H}_{port}$ described by the unitary operator $U$
\beq
U\, \ket{s}\ket{*}=\ket{\#}\ket{s}, \; \; \; \mbox{with} \; s=0,1,..., D-1
\label{definitia-sorter}
\eeq
is called {\bf quantum sorter}. The symbols $*$ and $\#$ are allowed to take the values 0, 1, 2, ..., $D-1$.

The purpose of the quantum sorter is to send the particle, which initially was in the state $\ket{s}$, to the output port-state $\ket{s}$:
\beqa
&&U\, \ket{\mbox{in-system-state}}\ket{\mbox{in-port-state}}\nonumber \\
&&=\ket{\mbox{out-system-state}}\ket{\mbox{out-port-state=in-system-state}}.
\eeqa
Remark: the symbol $*$ in Eq. (\ref{definitia-sorter}) may be unique for all $s=0,1,..., D-1$ or not, i.e. the particles could be incident on a unique input port or not. The only requirement imposed by the definition is that the operator $U$ to be unitary.

\section{Motivation}
A particular quantum sorter of a $D$-level quantum system was introduced in Ref. \cite{Ionicioiu} as a device with $D$ input ports and $D$ output ports, whose purpose is to sort the $D$ possible states of the quDit. The states of the quantum system exist in the Hilbert space  ${\cal H}_{system}$, while the states of the ports belong to the Hilbert space ${\cal H}_{port}$.
The quantum sorter of Ref. \cite{Ionicioiu} is defined by the unitary transformation acting on ${\cal H}_{system}\otimes {\cal H}_{port}$ as follows
\beq
\ket{s}\ket{k} \longrightarrow \ket{s}\ket{k \oplus s},\; \; \; \; s, k = 1, 2, \dots D,
\label{prima-def-single-sorter}
\eeq
By $\oplus $ we denote the addition {\it modulo D}. The $D$ output ports are used in order to collect the $D$ states of the quDit.

It may be worth noting that the quantum sorter given by Eq. (\ref{prima-def-single-sorter}) depends on the input port, i.e., there is a correspondence between the input and the output ports, the emerging one being influenced by the input. For example, let us consider that we inject two qutrits (three-level systems), which are in the same initial state $\ket{1}$, on the quantum sorter of Eq. (\ref{prima-def-single-sorter}). If one qutrit is incident on the mode $\ket{0}$ and the other one on the mode $\ket{1}$, then, according to Eq. (\ref{prima-def-single-sorter}), one obtains
\beqa
\ket{1}\ket{0} &\longrightarrow & \ket{1}\ket{1}, \nonumber \\
\ket{1}\ket{1} &\longrightarrow & \ket{1}\ket{2}.
\label{exemplu}
\eeqa
The two initial qutrits found in the same state $\ket{1}$ entering different input ports will exit through different output port-states (denoted by $\ket{1}$ and $\ket{2}$). But the purpose of the quantum sorter is to collect particles with the same property denoted by $s$ through the same output port, which should be unique for the particles found initial in the state $\ket{s}$.
This means that the quantum sorter of Eq. (\ref{prima-def-single-sorter}) fails in sorting many quantum systems, which are incident on different input ports.

The impossibility of a proper simultaneous selection of the quantum states of many particles by using the sorter (\ref{prima-def-single-sorter}) was our motivation for constructing the general quantum sorter which works for many particles incident on different input ports of a such device, as we have defined in Sect. 2.

\section{The two-input-port quantum polarization sorter of photons}

In this section we investigate the possibility of sorting two photons, which are incident on a polarizing beam splitter (PBS), according to their polarizations. The PBS is characterized by two-input modes and by two-output modes.
If one sends two $H$-polarized photons through the two different input ports of a PBS, then they will exit through different output ports. The PBS cannot be used as a quantum sorter when the two photons are injected through different input ports, because a proper quantum sorter would collect photons with identical polarization through the same output port.
This means that the PBS works as a quantum polarization sorter of two photons only in the case when the two photons are incident on the same input port.

In a setup of a quantum circuit, one expects that many particles are injected on a multi-port device.
This leads us to the motivation to find a scheme, which allows the simultaneous selection of many photons according to their initial polarizations, regardless of the input port. We impose the following requirement: the output mode $\ket{0}$ of this two-input two-output device will collect all the photons, which were initially $H$-polarized, while the output mode $\ket{1}$ collects the initially $V$-polarized photons, regardless of the final state of the photons.
In this way, by photocounting measurements, we obtain the information stored in the initial state of the photons.

\begin{figure}[thb]
\centering
\includegraphics[scale=0.64]{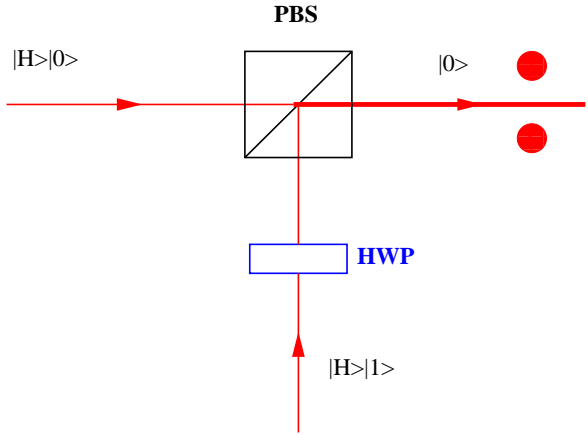}
\caption{The scheme of the two-input-port quantum polarization sorter of photons. The labels are as follows: PBS represents the polarizing beam splitter and HWP is the half-wave plate. The two $H$-polarized photons, which are incident on the input ports $\ket{0}$ and $\ket{1}$ of the quantum sorter, will exit through the output port $\ket{0}$.}
\label{fig-bs}
\end{figure}

Suppose that we insert a half-wave plate (HWP) in the arm of the input mode $\ket{1}$ of the PBS, as it is shown in Fig. \ref{fig-bs}. Again we sent two $H$-polarized photons on different input ports of this device. The action of the HWP is given by the $X$ gate, i.e. it flips the polarization state $\ket{H} \longleftrightarrow \ket{V}$ of the photons, which are incident on the input $\ket{1}$.
Then, the two photons exit through the same output-state, which is $\ket{0}$. The disadvantage of this device is that it flips the polarization of one of the photons, such that the emerging photons will be one $H$-polarized and the other one $V$-polarized.

The device shown in Fig. \ref{fig-bs} is called the two-input-port quantum polarization sorter of photons and consists of a PBS and a HWP placed in one arm of the PBS. The action of this quantum sorter of photons is described by the relations:

(i) for the incident $H$-polarized photons
\beqa
\ket{H}\ket{0} &\longrightarrow & \ket{H}\ket{0}  \label{H-pol-1}\\
\ket{H}\ket{1} &\longrightarrow & \ket{V}\ket{0}; \label{H-pol-2}
\eeqa

(ii) for the incident $V$-polarized photons
\beqa
\ket{V}\ket{0} &\longrightarrow & \ket{V}\ket{1} \label{V-pol-1} \\
\ket{V}\ket{1} &\longrightarrow & \ket{H}\ket{1}. \label{V-pol-2}
\eeqa

We can obtain the information stored in the initial states of the photons by measuring the photocounting distribution, i.e. from the click statistics of the detectors. In the example shown in Fig. \ref{fig-bs}, we can infer that two $H$-polarized photons were injected on the quantum sorter, since we obtain both photons measured by the detector in the output port-state $\ket{0}$.

If we denote $\ket{H}=\ket{0}$ and $\ket{V}=\ket{1}$, we can rewrite the above four expressions (\ref{H-pol-1}), (\ref{H-pol-2}), (\ref{V-pol-1}) and (\ref{V-pol-2}) in the compact form:
\beq
\ket{s}\ket{k} \longrightarrow \ket{s\ominus k}\ket{s},   \; \; s, k =0, 1,
\label{sort-fot}
\eeq
where $\ominus $ denotes the subtraction {\it modulo 2}.

\section{Classification of the quantum sorters of quDits}

The purpose of this section is to introduce and to make a classification of different kinds of quantum sorters, starting from the general definition (\ref{definitia-sorter}).

A quantum sorter would be perfect if it sends the particle, which is initially in the state $\ket{s}$, through the output port-state $\ket{s}$, regardless of the input port-state, while keeping the state of the particle unchanged.

\vspace{0.5cm}

{\bf Definition 2.} Consider two $D$-dimensional Hilbert spaces ${\cal H}_{system}$ and ${\cal H}_{port}$, the first one associated to a quDit system, while the second one to the states of the $D$ ports.
A device acting in the Hilbert space ${\cal H}_{system}\otimes {\cal H}_{port}$, that performs the unitary transformation
\beq
\ket{s}\ket{k} \longrightarrow \ket{s}\ket{s}, \; \; \; \mbox{with} \; s, k=0,1,..., D-1
\label{perfect-m-sorter}
\eeq
is called {\bf the perfect multi-input-port quantum sorter}.

An equivalent way of writing Eq. (\ref{perfect-m-sorter}) is as follows:
\beqa
&&U\, \ket{\mbox{in-system-state}}\ket{\mbox{in-port-state}}\nonumber \\
&&=\ket{\mbox{out-system-state=in-system-state}}\ket{\mbox{out-port-state=in-system-state}}.\nonumber
\eeqa

In the following, we prove that the perfect multi-input-port quantum sorter is forbidden.
Suppose that we sent two particles found in the same state $\ket{s}$ on this device, one through the input port-state $\ket{m}$, while the second one through $\ket{n}$:
\beqa
\ket{s}\ket{m} &\longrightarrow & \ket{s}\ket{s}; \nonumber \\
\ket{s}\ket{n} &\longrightarrow & \ket{s}\ket{s}, \; \; \mbox{with} \; m\ne n.
\label{no-univ}
\eeqa
The proof is straightforward, since there is no unitary operator satisfying both Eqs. (\ref{no-univ}).

\vspace{0.5cm}

{\bf Definition 3.}  Consider two $D$-dimensional Hilbert spaces ${\cal H}_{system}$ and ${\cal H}_{port}$, the first one associated with a quDit system, while the second one to the states of the $D$ ports.
A device that performs the following unitary transformation acting in the Hilbert space ${\cal H}_{system}\otimes {\cal H}_{port}$
\beq
U_{SQS}\, \ket{s}\ket{k} = \ket{s}\ket{s \oplus k}= \left\{ \begin{array}{lcl}
\ket{s}\ket{s}, & \mbox{if} & k=0 \\
\ket{s}\ket{j}, &\mbox{with} \; j=s \oplus k\ne s, \; \mbox{if} & k\ne 0,
\end{array} \right.
\label{single-sorter}
\eeq
with $k=0$ and $s$ = 0, 1,..., $D-1$, is called {\bf the single-input-port quantum sorter (SQS)}. The incident particles must be injected only on the input port-state $\ket{0}$.

The action of the quantum sorter of Eq. (\ref{single-sorter}) can be rewritten as:
\beqa
&&U_{SQS}\, \ket{\mbox{in-system-state}}\ket{\mbox{in-port-state=0}}\nonumber \\
&&=\ket{\mbox{out-system-state=in-system-state}}\ket{\mbox{out-port-state=in-system-state}}. \nonumber
\eeqa
The unitary operator $U_{SQS}$ satisfies the condition of the general quantum sorter given by Eq. (\ref{definitia-sorter}) only for the input port $s=0$; therefore it is called single-input-port quantum sorter. The device of Eq. (\ref{single-sorter}) was introduced in Ref. \cite{Ionicioiu} and it was analyzed in Sect. 3, where we have emphasized that this quantum sorter fails in selecting many particles, which are incident on different input ports. In Ref. \cite{Ionicioiu}, it was proved that:
\beq
U_{SQS}=C(X_D)=(I\otimes F^\dagger )\, C(Z_D)\, (I\otimes F),
\label{expr-single}
\eeq
where $F$ is the quantum Fourier transform (QFT) defined by Eq. (\ref{fourier}), while $C(X_D)$ and $C(Z_D)$ are the controlled-$X_D$ gate and controlled-$Z_D$ gate, respectively, given by Eqs. (\ref{xd}) and (\ref{zd}).

Since the perfect multi-input-port quantum sorter of many quDits is forbidden, we try to construct a quantum sorter, which may change the state of the particles, while the particles exit through the proper output port dependent on the initial state. In the following, we give the definition of a such device.

\vspace{0.5cm}

{\bf Definition 4.} Consider two $D$-dimensional Hilbert spaces ${\cal H}_{system}$ and ${\cal H}_{port}$, the first one associated with a quDit system, while the second one to the states of the $D$ ports.
A device acting in the Hilbert space ${\cal H}_{system}\otimes {\cal H}_{port}$, that performs the unitary transformation
\beq
U_{MQS}\, \ket{s}\ket{k} = \ket{s \ominus k}\ket{s}, \; \; \; \mbox{with} \; s, k=0,1,..., D-1
\label{m-sorter}
\eeq
is called {\bf the multi-input-port quantum sorter (MQS)}.

This device sends the particle, initially found in the state $\ket{s}$, through the output state $\ket{s}$, regardless of the input port state, while the state of the particle is changed to $\ket{s \ominus k}$, with $\ominus $ the subtraction {\it modulo D}:
\beqa
&&U_{MQS}\, \ket{\mbox{in-system-state}}\ket{\mbox{in-port-state}}\nonumber \\
&&=\ket{\mbox{out-system-state=in-system-state}-\mbox{in-port-state}}\nonumber \\
&&\otimes\ket{\mbox{out-port-state=in-system-state}}.\nonumber
\eeqa

Or in other words, it performs an approximate sorter action of many incident particles, where the task is only to count the number of the particles which were initial in the state $\ket{s}$, with no interest in the state of the output particles. Therefore, we can extract the information stored in the initial states  from the click statistics of the detectors. The MQS is useful in the case when high speed of quantum state sorting is required, since the sorting efficiency is increased by a factor of $D$.

In Table \ref{definitii}, we present the classification of the quantum sorters, by emphasizing on which input port-states they work, as well as the equation which defines each sorter.

\begin{table}[h]
\caption{The classification of quantum sorters. The incident particle is found in the state $\ket{s}$. The word "YES" means that the condition of the general quantum sorter (\ref{definitia-sorter}) is satisfied, i.e. the particle exits through the output port-state $\ket{s}$. The word "NO" has the meaning that the particle emerges through the output port-state $\ket{j}$, with $j\ne s$. We have proved that the perfect quantum sorter cannot be constructed.}
\begin{center}
\begin{tabular}{|c|c|c|c|}
\hline
Quantum & \multicolumn{2}{c|}{Input port-states}& Equation which\\
\cline{2-3}
 Sorter          & $\ket{0}$ & $\ket{1}$, $\ket{2}$,..., $\ket{D-1}$ & defines the sorter \\
\hline \hline
Perfect & YES & YES & Eq. (\ref{perfect-m-sorter})\\
\hline
Single-input-port& YES & NO & Eq. (\ref{single-sorter})  \\
\hline
Multi-input-port& YES & YES & Eq. (\ref{m-sorter})  \\
\hline
\end{tabular}
\end{center}
\label{definitii}
\end{table}

Let us return now to the discussion regarding the PBS by using the new definitions introduced above.
Since the PBS works as a quantum sorter only if all the photons are injected on the same input mode, it represents the single-input-port polarization sorter of the photons according to Definition 3. The device presented in Fig. \ref{fig-bs} in Sect. 4  constructed with the help of a PBS and a half-wave plate is an example of two-input-port quantum sorter of qubits due to Eq. (\ref{sort-fot}) and Definition 4.

Let us discuss again the example of Eq. (\ref{exemplu}) (presented in Sec. 3), by considering now the scenario when the two qutrits are incident on the multi-input-port quantum sorter given by Eq. (\ref{m-sorter}):
\beqa
\ket{1}\ket{0} &\longrightarrow & \ket{1}\ket{1}, \nonumber \\
\ket{1}\ket{1} &\longrightarrow & \ket{0}\ket{1}.
\eeqa
Both particles initially found in the state $\ket{1}$ incident on different modes will exit through the output port $\ket{1}$, as one requires for a quantum sorter.

\section{The quantum circuit of the multi-input-port quantum sorter}

In the following, we find an equivalent expression of the unitary operator $U_{MQS}$ in terms of the well-known quantum gates of quDits, as well as the quantum circuit associated to the multi-input-port quantum sorter.

\vspace{0.5cm}

{\bf Theorem.} {\it The unitary operator that satisfies condition (\ref{m-sorter}) of the multi-input-port quantum sorter is given by}
\beq
U_{MQS}=U_{SQS}\, \left( (X_D^\dagger )^k\otimes I\right),
\label{math-gen-sort}
\eeq
{\it where $k$ is the input port of the incident particle, while the expression of $U_{SQS}$ is given by Eq. (\ref{expr-single})}.

\vspace{0.3cm}

{\bf Proof.}
By using the action of the operator $X_D^\dagger $ according to Eq. (\ref{xd-kori}) of Appendix A, one obtains
\beq
(X_D^\dagger )^k\otimes I\, \ket{s}\ket{k} = \ket{s \ominus k}\ket{k}.
\eeq
Further, by applying the single-input-port quantum sorter (\ref{single-sorter}), we have
\beq
U_{SQS}\, \ket{s \ominus k}\ket{k}= \ket{s \ominus k}\ket{s},
\eeq
i.e. the desired outcome. $\Box $

\vspace{0.3cm}

It may be worth noting that in the particular case $D=2$, one has $X^\dagger =X$ and this property was used in Sect. 4, when we have discussed the two-input-port quantum polarization sorter for photons (see also Fig. \ref{fig-bs}).

\begin{figure*}[thb]
\centering
\includegraphics[scale=0.5]{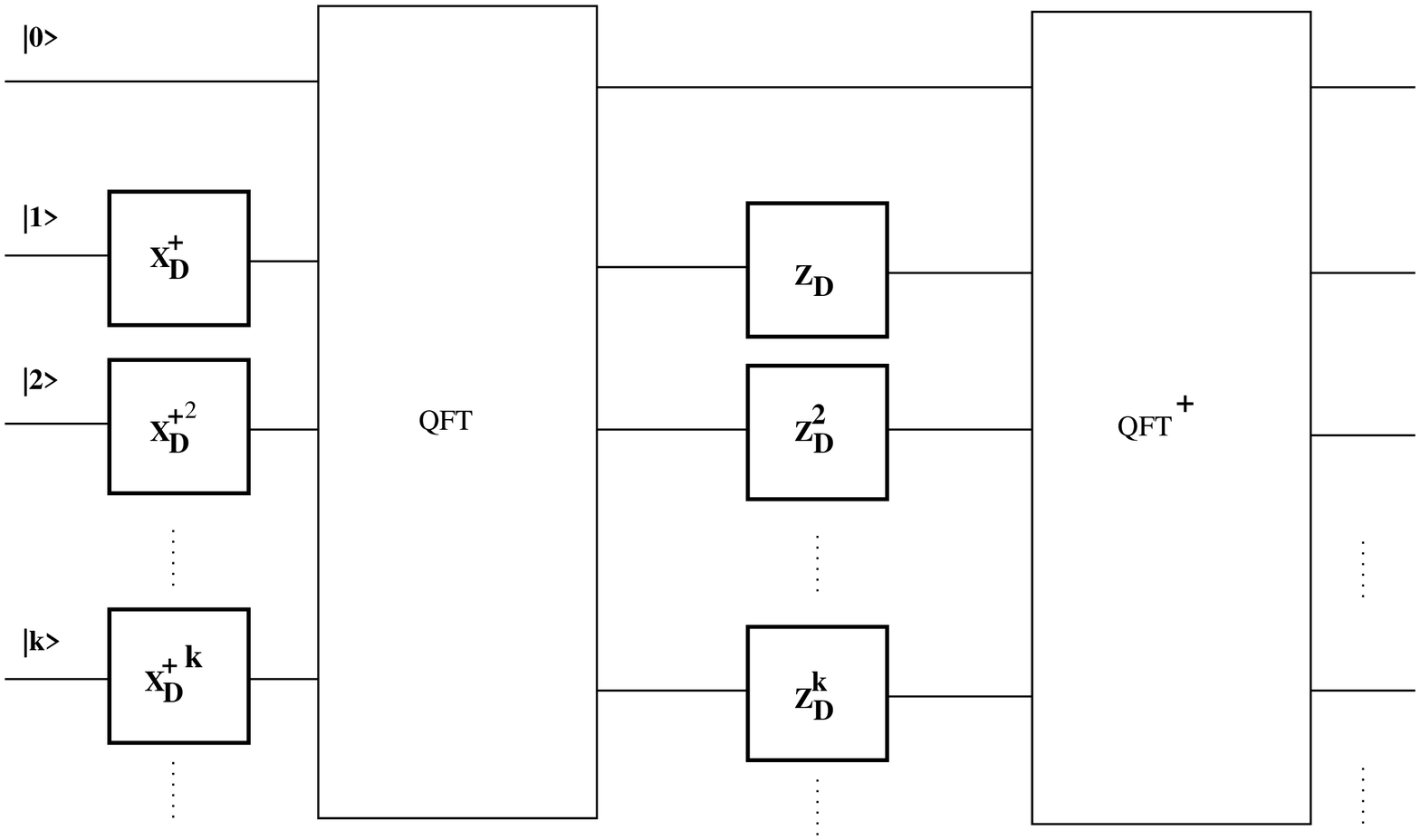}
\caption{The quantum circuit of the multi-input-port quantum sorter of quDits. The input and the output port-states are indexed by $\ket{0}$, $\ket{1}$,..., $\ket{D-1}$. A particle found in the quantum state $\ket{s}$, which is injected to the input port-state $\ket{k}$, with $k$ arbitrary, will exit through the output port-state $\ket{s}$, while its state is modified to $\ket{s\ominus k}$, according to Eq. (\ref{m-sorter}) and the Theorem described by Eq. (\ref{math-gen-sort}). This device sorts simultaneously many particles found in the states $\ket{s_1}$, $\ket{s_2}$,... regardless of the input port-state. }
\label{fig-multi-sort}
\end{figure*}

The quantum circuit of the multi-input-port quantum sorter is depicted in Fig. \ref{fig-multi-sort}. Firstly, the state of a particle injected on the input-state $\ket{k}$ is modified by the $(X_D^\dagger )^k$ gate. Secondly, according to the Theorem described by Eq. (\ref{math-gen-sort}), one uses the scheme of the SQS.  In Ref. \cite{Ionicioiu}, it was shown that the SQS consists of a QFT, which has the role of coupling all the modes with the same probability $1/D$, followed by a path-dependent $Z_D^k$ gate and further an inverse of the QFT.

The realization of experimental high-dimensional gates for any degrees of freedom of the quantum systems is a challenging task and still an open problem. Schemes for the implementation of the $Z_D$ gate defined by Eq. (\ref{z-gate}) for photons carrying OAM were proposed in Ref. \cite{Boyd-2013} by using a Dove prism, as well as for the $Z_D^k$ gate were found by employing also the Dove prisms rotated by a specific angle \cite{Ionicioiu}. The generalized Pauli gate $X_D$ of Eq. (\ref{x-gate}) seems to be more difficult to be generated in the lab. Recently it was experimentally demonstrated the construction of the $X_D$ gate and all of its integer powers on single photons with OAM for $D=4$ in Ref. \cite{Zeilinger-PRL-2017}.

\begin{figure*}[thb]
\centering
\includegraphics[scale=0.5]{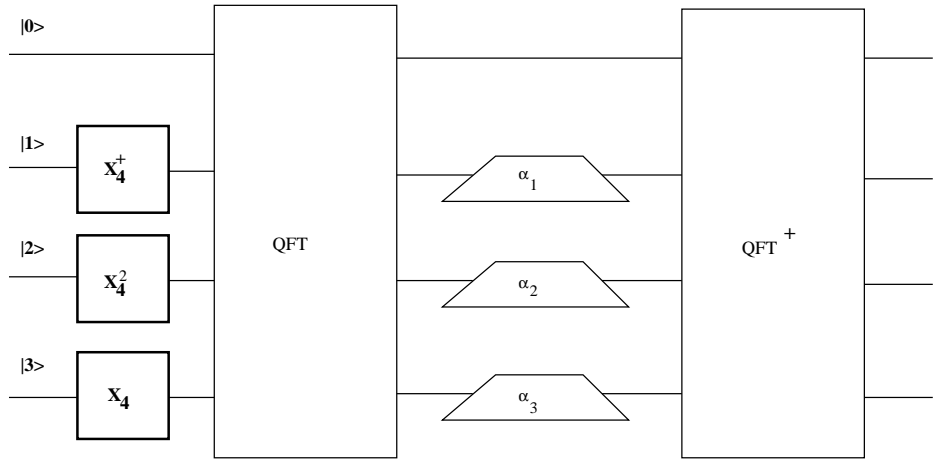}
\caption{The four-input-port quantum sorter of four-dimensional OAM of photons. The input and the output ports are indexed by $\ket{0}$, $\ket{1}$, $\ket{2}$, $\ket{3}$. The schematic diagrams of the gates $X_4^\dagger $, $X_4^2$, and $X_4$ are given in Fig. 1 of Ref. \cite{Zeilinger-PRL-2017}, being constructed with the help of a  Mach-Zehnder interferometer, two parity splitters, and a spiral phase plate.  QFT is the quantum Fourier transform. Further, in the modes $\ket{1}$, $\ket{2}$, $\ket{3}$ is placed a Dove prism rotated by the angle $\alpha_k$ of Eq. (\ref{alfa}). This device sorts simultaneously many photons carrying OAM for $D=4$, which are injected on any input port. }
\label{fig-oam-sort}
\end{figure*}

The physical implementation of the four-input-port quantum sorter for photons carrying four-dimensional OAM can be realized as it is shown in Fig. \ref{fig-oam-sort}. On the input mode $\ket{1}$, one applies the gate $X_4^\dagger$, whose scheme is shown in Fig. 1(c) of Ref. \cite{Zeilinger-PRL-2017}, consisting of a Mach-Zehnder interferometer, two parity splitters, and a spiral phase plate placed in the output mode. By using the fact that $(X_4^\dagger )^2=X_4^2$, we place in the input mode $\ket{2}$ the gate $X_4^2$, whose representation is given in Fig. 1(b) of Ref. \cite{Zeilinger-PRL-2017}, i.e. a Mach-Zehnder interferometer, two parity splitters, and a spiral phase plate placed in the even arm of the interferometer. On the last input mode, it is sufficient to use the gate $X_4$ due to the fact $(X_4^\dagger )^3=X_4$ - the diagram of this gate is drawn in Fig. 1(a) of Ref. \cite{Zeilinger-PRL-2017}: again a  Mach-Zehnder interferometer, two parity splitters, and a spiral phase plate situated in the input mode. Further, a quantum Fourier gate is applied, followed by the action of a Dove prism placed in the modes $k$ = 1, 2, 3 rotated with the angles $\alpha_1$, $\alpha_2$, $\alpha_3$, respectively:
\beq
\alpha_k = k\, \frac{\pi}{D}.
\label{alfa}
\eeq
Finally, the inverse of the QFT is used on all the modes.

\section{Conclusions}

In conclusion, in this paper we have introduced the definition of the general quantum sorter of quDits given by Eq. (\ref{definitia-sorter}). With the help of this definition, we have made a classification of the quantum sorters and have
shown the impossibility of construction of the perfect multi-input-port quantum sorter, which simultaneously selects many quDits, while keeping their states unmodified. This fact led us to the proposal of the multi-input-port quantum sorter that performs an approximate selection. Importantly, in contrast to the SQS, which works properly only in the case when all the particles are incident on the input port-state $\ket{0}$, our MQS has the ability to simultaneously collect many particles, which initially were in the same state, to the same output port-state. Therefore, the speed of quantum state sorting is increased by a factor of $D$. The price to be paid is the modification of the state of the output particles.

Our scheme of the MQS may be suited to further applications to quantum information processes, because it performs the sorting of many multi-level particles by measuring the number of the particles in each output port. This means that from the click statistics of the detectors one can reconstruct the information stored in the initial states of the particles.

\begin{acknowledgements}
I wish to thank one of the two anonymous referees for his/her valuable suggestions, which led to an improved version of this paper. This work was supported by the funding agency CNCS-UEFISCDI of the Romanian Ministry of Research and Innovation through Grant PN-III- P4-ID-PCE-2016-0794 within PNCDI III.
\end{acknowledgements}

\appendix
\section{Quantum gates acting on quDits}

In this appendix, we provide a brief review of the definitions of the most important quantum gates acting on one and two quDits, such as the generalized Pauli operators, the quantum Fourier transform, and the controlled-U gate.

If $\{ \ket{0}$, $\ket{1}$,..., $\ket{D-1}\}$ is the basis of the $D$-dimensional Hilbert space of the quDit, then the generalized Pauli operators are defined as \cite{Gottesman}, \cite{Zeilinger-light-2018}
\beqa
X_D&=&\sum_{j=0}^{D-1}\, \proj{j\oplus 1}{j}; \label{x-gate} \\
Z_D&=&\sum_{j=0}^{D-1}\, \omega^j\, \proj{j}{j},\label{z-gate}
\eeqa
where addition is taken {\it modulo} $D$ and $\omega =\exp (2\pi i/D)$ is the $D$th root of unity. The particular case $D=2$ corresponds to the regular Pauli operators.

After applying the inverse of $X_D$ $k$ times, one obtains
\beq
\left( X_D^\dagger \right) ^k \ket{s}=\ket{s\ominus k}.
\label{xd-kori}
\eeq

The quantum Fourier transform (QFT) has the expression \cite{Nielsenbook}:
\beq
F\ket{k}=\frac{1}{\sqrt{D}}\, \sum_{j=0}^{D-1}\, \omega^{k\, j}\, \ket{j}.
\label{fourier}
\eeq
The controlled-$U$ gate acting on two quDits is defined as \cite{Ionicioiu}:
\[
C(U)\, \ket{s}\ket{k}=\ket{s}\, U^s \ket{k},
\]
which in the particular cases of $X_D$ and $Z_D$ gates reads:
\beqa
C(X_D)\ket{s}\ket{k}&=&\ket{s}\ket{k\oplus s} \label{xd} \\
C(Z_D)\ket{s}\ket{k}&=&\omega^{s\, k}\ket{s}\ket{k}\label{zd}.
\eeqa

\end{document}